\documentclass[12pt,english,letterpaper]{article}
\pdfoutput=1

\usepackage{natbib} 
\bibpunct{(}{)}{;}{a}{}{,}

\usepackage[english]{babel}

\usepackage{amsmath}

\usepackage{mathptmx}

\usepackage[T1]{fontenc} 
\usepackage[utf8]{inputenc} 

\usepackage{amssymb}
\usepackage{mathtools}

\usepackage{graphicx}

\usepackage{microtype} 
\usepackage{booktabs}  
\usepackage{array}     

\usepackage{titlesec}
\titleformat{\section}[block]{}{\thesection.}{5pt}{\bf\normalsize}
\titleformat{\subsection}[block]{}{\thesubsection.}{5pt}{\normalsize\emph}
\titleformat{\subsubsection}[block]{}{\thesubsubsection.}{5pt}{}

\usepackage[width=6in]{geometry} 

\bibpunct{(}{)}{;}{a}{}{,}

\newcommand{\var}{{\rm Var}}

\title{Deviation test construction and power comparison for marked spatial point patterns}
\author{Mari Myllym{\"a}ki$^{1}$, Pavel Grabarnik$^{2}$, Henri Seijo$^{1}$, Dietrich Stoyan$^{3}$ \\
$^{1}$ Department of Biomedical Engineering and Computational Science, \\
Aalto University School of Science, Finland, \\
$^{2}$ Institute of Physico-Chemical and Biological Problems in Soil Science, \\
the Russian Academy of Sciences, Russia, \\
$^{3}$ Institut f\"{u}r Stochastik, TU Bergakademie Freiberg, Germany}
\date{}

\begin{document}

\maketitle

\begin{abstract}
The deviation test belong to core tools in point process statistics,
where hypotheses are typically tested considering
differences between an empirical summary function and its
expectation under the null hypothesis, which depend on a distance
variable $r$. This test is a classical device to overcome the
multiple comparison problem which appears
since the functional differences have to be considered for a range
of distances $r$ simultaneously. The test has three basic
ingredients: (i) choice of a suitable summary function, (ii)
transformation of the summary function or scaling of the
differences, and (iii) calculation of a global deviation measure.
We consider in detail the construction of such tests both for
stationary and finite point processes and show by two toy examples
and a simulation study for the case of the random labelling
hypothesis that the points (i) and (ii) have great influence on the
power of the tests.
\end{abstract}

\noindent \emph{Key words}: deviation test; marked point process;
marking model; mark-weighted $K$-function; Monte Carlo test;
multiple comparison; random labelling; simulation study

\section{Introduction}

Testing statistical hypotheses is an important step in building
statistical models. Often it is checked whether the data deviate
significantly from a null model. In point process statistics,
typical null models are complete spatial randomness (CSR),
independent marking or some fitted model. Unlike in classical
statistics, where null models are typically represented by a single
hypothesis, the hypotheses in spatial statistics have a spatial
dimension and therefore a multiple character. Usually a summary
function $S(r)$ is employed in the test, where $r$ is a distance
variable. A typical example is Ripley's $K$-function.

The tests are based on the differences of empirical and theoretical
values of $S(r)$, which are called ``residuals'' in the following. A
problem is how to handle the residuals for different values of $r$.
One possibility is construction of envelopes around the theoretical
summary function and to look if the empirical summary function is
completely between the envelopes. This very popular method, which
goes back to \citet{Ripley1977}, has a difficult point: to
guarantee a given significance level and to determine $p$-values,
see the discussion in \citet{LoosmoreFord2006} and
\citet{GrabarnikEtal2011}. An alternative approach was proposed by
\citet{Diggle1979}, who introduced statistics which compress
information from the residuals for intervals of $r$-values to a
scalar. This approach has analogues in classical statistics, namely
the Kolmogorov-Smirnov and von Mises tests. In the present paper
tests in Diggle's spirit are called {\em deviation tests}.

Though Diggle's procedure is accepted as a standard in spatial point
process statistics, to our knowledge there are no studies
which explore its properties systematically. Several power
comparisons for different forms of deviation tests have been
reported \citep[e.g.][]{Ripley1979, Gignouxetal1999,
ThonnesLieshout1999, BaddeleyEtal2000b, GrabarnikChiu2002,
HoChiu2006, HoChiu2009}, but these investigations concern only
specific issues. In the present paper, we
consider the construction of deviation tests in more detail and
systematically and give general recommendations for their use, for
stationary as well as for finite point processes.

Recall that a classical deviation test in point process statistics
is based on a summary function $S(r)$ for the null model and its
unbiased estimator $\hat{S}(r)$. If there would be {\em a priori} a
value of distance $r$ which is of main interest, then one could
proceed as in classical tests by comparing the empirical
value $\hat{S}(r)$ with $S(r)$ for this $r$, i.e.\ to consider the
residual $\hat{S}(r)-S(r)$. However, since usually such a single
special distance $r$ is not given, one would like to consider the
residuals simultaneously for all distances $r$ in some interval
$I=[r_{\text{min}}, r_{\text{max}}]$. Thus, one is confronted with a
situation typical for multiple hypothesis testing (or multiple
comparisons), see \citet{BretzEtal2010}. A deviation test resolves
the multiple hypothesis testing problem by summarizing the residuals
for all $r$ in $I$ into a single number by some deviation measure,
e.g.\ the maximum absolute residual in $I$.

A summary function frequently used in point process statistics for
stationary processes and in deviation tests is Ripley's $K$-function
\citep{Ripley1976, Ripley1977}. Since \citet{Besag1977} found that,
under CSR, the $L$-function resulting from the transformation
$\hat{K}(r) \rightarrow \sqrt{\hat{K}(r)/\pi}$ has a variance which
is approximately constant over the distances $r$, the $L$-function
became popular. Consequently, it is empirically well-accepted that a deviation
test based on the $L$-function is ``better'' than a test based on
the $K$-function. Variance-stabilising transformations are available
also in other cases, see e.g.\ \citet{SchladitzBaddeley2000} and
\citet{GrabarnikChiu2002}, whereas many summary functions such as
the nearest neighbour distribution function ($D$- or $G$-function)
and $J$-function \citep{LieshoutBaddeley1996} are employed without transformations.

We show that it is useful to use transformations and additional
scalings of the residuals, particularly relative to their variances,
in order to uniform the contributions of residuals for different
distances $r$.

Besides transformations and scaling also other elements of deviation
tests have influence on its power. These are the basic choice of the
summary function $S(r)$, the interval of distances $I$ and
the deviation measure.
We explore the role of these elements through a
simulation study for an important test problem for marked point
processes: checking the random labelling hypothesis, which says that
the marks of a marked point pattern are independent. Note that this
particular case contains all elements of a deviation test in point
process statistics, for the stationary as well as for the finite
case.

The rest of the paper is organized as follows. Section
\ref{sec:background} explains the fundamentals of deviation tests in
point process statistics, in a generalised form applicable also for
non-stationary and finite processes
and biased estimators. Section \ref{sec:construction} then discusses the construction of
deviation tests in detail, from the point of view of multiple
testing. Section \ref{sec:simulation_study_setup} describes the
design for the simulation study, the results of which are reported
in Section \ref{sec:simstudy}. Section \ref{sec:discussion}
discusses the results and ends with recommendations for practical
work.

\section{Preliminaries on point process statistics}\label{sec:background}

The symbols $N=\{x_i\}$ and $N_m=\{[x_i;m(x_i)]\}$ denote a point
process and a marked point process, respectively. The $x_i$ are the
points, while $m(x_i)$ is the mark of point $x_i$. In the present
paper planar point processes are considered, but the main ideas hold
true also for point processes in $\mathbb{R}^d$ for $d\geq 2$.

If $N$ and $N_m$ are stationary, they have an intensity which is
denoted by $\lambda$. The mark distribution function is denoted by
$F_M(m)$. Its mean and variance are $\mu_m$ and $\sigma^2_m$.

The statistical analysis is based on observations in a window $W$,
which is a compact convex subset of $\mathbb{R}^d$. In the planar
case the window is often a rectangle.

When distributional hypotheses for point processes have to be
tested, deviation tests as suggested by \citet{Diggle1979} are a
popular tool. We consider these tests here in a generalised form.
Such a test is based on a test function $T(r)$ that characterises
in some way the spatial arrangement of the points and/or marks in
the window $W$. There are many possibilities for such functions. In
the classical case, a common choice for $T(r)$ is an unbiased
estimator $\hat{S}(r)$ of some summary function $S(r)$,
see e.g.\ \citet{Cressie1993}, \citet{Diggle2003} and
\citet{IllianEtal2008} for examples.
Popular functions for stationary processes
in the case without marks are Ripley's $K$-function, the nearest
neighbour distance distribution function ($G$-function)
and the empty space function/spherical contact distribution ($F$-function).
For stationary marked point processes, various mark
correlation functions including the mark-weighted $K$-functions are
available.
For non-stationary or finite processes analogues of
stationary-case characteristics can be used.

Since deviation tests are in essence Monte Carlo tests, one needs to
be able to generate spatial point patterns for the tested null model
in $W$ and to calculate $T(r)$ for data and each simulated pattern.
The function for data is denoted below by $T_1(r)$ and for $s$
simulations the corresponding functions are $T_i(r)$ for
$i=2,\ldots, s+1$.
These functions are then compared with the expectation
$T_0(r)$ of $T(r)$ for the null model in $W$.
For this a global deviation measure $u_i$ is used that summarises
the discrepancy between $T_i(r)$ and $T_0(r)$ into a single number
$u_i$ for all $r \in I$.

The deviation measure is calculated for the data ($u_{1}$) and for
the $s$ simulated patterns from the null model
($u_2,\dots,u_{s+1}$). The rank of $u_1$ among the $u_i$ is the
basis of the Monte Carlo test \citep{Barnard1963, BesagDiggle1977}.

There are various ways to obtain $T_0(r)$. The classical case is
that of a stationary point process, an unbiased estimator
$\hat{S}(r)$ of a summary function $S(r)$ and the known form of
$S(r)$ for the null model. Then it is simply $T(r) = \hat{S}(r)$ and
$T_0(r)=S(r)$. Usually edge-corrected estimators are then needed.
Also in some other cases, the expectation $T_0(r)$ is analytically
known, as for the example considered in Section
\ref{sec:simulation_study_setup}. Otherwise, $T_0(r)$ has to be
determined statistically based on simulations of the null model in
the window $W$.

A simple estimator of $T_0(r)$ is the mean of $m$ functions
$T_j(r)$ obtained from another independent set of $m$ simulations of the null
model in $W$. In order to save computing time,
the same samples can be used for determining the $u_i$ and
$T_0(r)$.
\citet[p.\,14]{Diggle2003} suggested to use for each simulation $i$,
$i=1,\dots,s+1$ ($i=1$ is data), its own mean value
$\bar T_i(r)  =  \sum_{j=1, j\neq i}^{s+1}  T_j(r) / s$.
In this case the statistics $u_i$ are exchangeable and,
under the null hypothesis, all rankings of $u_1$ among the $u_i$ are
equiprobable as in the case where $T_0$ is obtained analytically or
from another set of simulations.

\section{Deviation test construction}\label{sec:construction}

This section discusses in detail how the global deviation test
statistic $u$ is constructed from a test function $T(r)$
of a pattern observed in a window $W$
and its expectation $T_0(r)$ under the null hypothesis.

\subsection{Raw residuals and deviation measures}
The raw residual is simply
\begin{equation}\label{rawresidual}
d(r) = T(r) - T_{0}(r) \quad \text{for } r \ge 0.
\end{equation}
All raw residuals for $r \in I = [r_{\text{min}},r_{\text{max}}]$
are summarised into a global deviation measure $U$.
Examples are the supremum deviation measure
\begin{equation}\label{Dinfty}
U_{\infty}= \sup_{r \in I} |d(r)| = \sup_{r \in I} \Big| T(r) - T_{0}(r) \Big|
\end{equation}
and the integral deviation measure
\begin{equation}\label{DL2}
U_{L^2} =  \int_I \left(d(r)\right)^2 \text{d}r = \int_I \left( T(r) - T_{0}(r )\right)^2 \text{d}r.
\end{equation}
Both measures are used in the present paper, both in a discretised form.

\subsection{Residuals of transformed summary functions}

Probably in many cases it makes sense to transform summary functions
in the context of deviation tests. While for interpretation and
description the traditional summary characteristics should be used,
for tests modifications or transformations may behave better. Use of
a transformation function $h:\mathbb{R}\rightarrow\mathbb{R}$ leads
to residuals
$d_{h}(r) =  h(T(r)) - h(T_0(r))$.

Besides the square root transformation used in the context of the
$L$-function, a further example is the transformation $h(\cdot) =
\sqrt[4]{\cdot}$ used by \citet{SchladitzBaddeley2000} and
\citet{GrabarnikChiu2002}, e.g., for a third order analogue of
Ripley's $K$ function for planar processes. For $F$- and
$G$-functions the Aitkin-Clayton variance stabilising
transformation $h(\cdot)=\arcsin(\sqrt{1-\cdot})$
may be useful \citep{AitkinClayton1980}.

The present paper uses the well-established square root
transformation $h(\cdot)=\sqrt{\cdot/\pi}$, also in the context of
marked point processes.
Note that, in the general case of $\mathbb{R}^d$, the
$L$-function is defined by the transformation
$h(\cdot)=\sqrt[d]{\cdot/b_d}$ of the $K$-function, where $b_d$ is
the volume of a $d$-dimensional unit ball (for $d=2$, $b_d=\pi$),
see \citet{IllianEtal2008}.

\subsection{Scaled residuals}

Sometimes the variation of raw residuals differs clearly for the
$r$-values in the interval $I$, and therefore the residuals
contribute differently to the global deviation measure $U$.
The contributions can by made more equal by making the distribution of residuals
more uniform in $I$. This can be carried out by weighting the raw
residuals by weights $w(r)$ which depend on the distribution of
$T(r)$ under the null hypothesis, and by working with scaled
residuals
$d_{w}(r) = w(r) d(r)$.

Two natural choices are {\em studentised} scaling
\begin{equation}\label{d_var}
d_{\text{st}}(r)= \frac{d(r)}{\sqrt{\var_{0}(T(r))}},
\end{equation}
and {\em quantile} scaling
\begin{equation}\label{d_env}
 d_{\text{q}}(r)= \frac{d(r)}{\overline T(r) - \underline T(r)},
\end{equation}
where $\var_{0}(T(r))$ denotes the variance of $T(r)$ under the null
model, and $\overline T(r)$ and $\underline T(r)$ are the $r$-wise
2.5$\%$-upper and -lower quantiles of the distribution of $T(r)$
under $H_0$. These weights are typically not available analytically
but can be easily determined by simulation similarly as $T_0(r)$.
\citet{BaddeleyEtal2000b} applied the studentised scaling
\eqref{d_var} to the $J$-function, while the quantile scaling
\eqref{d_env} was used by \citet{MollerBerthelsen2012} in the
context of the (centred) $L$-function.

The following toy example aims to show why often scaling improves
the power in multiple tests, but sometimes not, and that it is
valuable to know the variability of residuals. The index $i$ in the
example corresponds to distance $r$ in point process statistics.

{\em Toy example 1.} Consider three normally distributed random
variables $X_1$, $X_2$ and $X_3$ with known variances $\sigma_i^2$,
$i=1,2,3$. The null hypothesis is that all three $X_i$ have mean
zero. The alternative model has a shift at $i=3$: $X_i \sim N(\mu_i,
\sigma_i^2)$ with $\mu_1=\mu_2=0$ and $\mu_3>0$.

Apply a deviation test to check the null hypothesis, with the global
deviation measures $U_{\infty} = \max_{i} |X_i|$ and $U_{\infty,w} =
\max_{i} (|X_i|/\sigma_i)$ based on raw and scaled residuals, \eqref{rawresidual} and \eqref{d_var}, respectively.
The power of these tests can be calculated analytically, see the Appendix.

Figure \ref{Fig_toyexample1} shows power curves for the significance level
$\alpha=0.05$ as a function of $\mu_3$ for two
different cases where (a) $\sigma_3$ is smaller than $\sigma_1$ and $\sigma_2$
and (b) $\sigma_3$ is larger than $\sigma_1$ and $\sigma_2$
In case (a) the scaled test with deviation measure $U_{\infty,w}$ is superior in
power, since the maximum of $|X_i|$
is most often reached at $X_1$ or $X_2$
and the unscaled test has therefore difficulties to observe the
deviation of $\mu_3$ from zero. In contrast, in case (b) scaling is
counter productive. The reason is that now the largest variance
occurs for the interesting index $i=3$, and the unscaled test resembles
a single hypothesis test.

\begin{figure}[h!]
\centering
\includegraphics{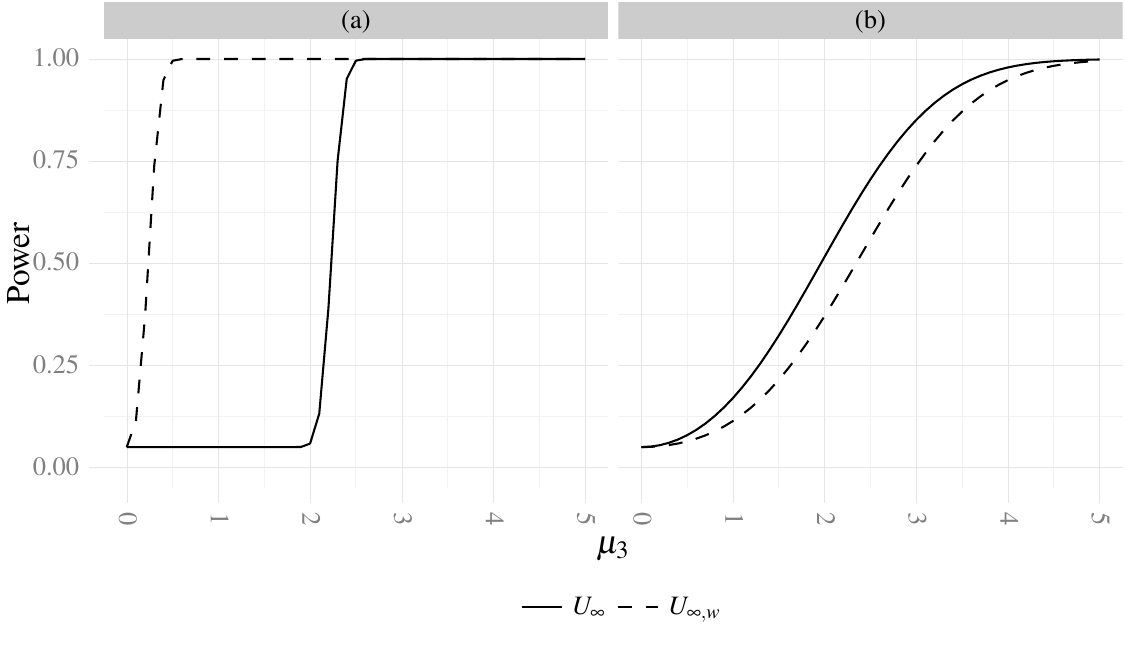}
\caption{Power curves as a function of $\mu_3$ for the alternative
model $X_i \sim N(\mu_i, \sigma_i^2)$ with $\mu_1=\mu_2=0$ and
$\mu_3>0$ against the null hypothesis that $\mu_i = 0$ for all
$i=1,2,3$. The standard deviations are (a) $\sigma_1=\sigma_2=1$ and
$\sigma_3=0.1$, (b) $\sigma_1=\sigma_2=0.1$ and $\sigma_3=1$. The
lines represent the tests based on the deviation measures
$U_{\infty}$ (unscaled test) and $U_{\infty,w}$ (scaled test).
See the text for details.}\label{Fig_toyexample1}
\end{figure}

\subsection{Asymmetric distribution of residuals}

Sometimes the distribution of the residuals (under the null model)
may be clearly asymmetric around $T_0(r)$.
Then transformations making the distribution more symmetric are helpful.
A simple directional scaling which treats negative and positive residuals differently, is
\begin{equation}\label{d_envdir}
 d_{\text{qdir}}(r)= \mathbf{1}(d(r)\geq 0)\frac{d(r)}{ |\overline T(r) - T_0(r)| }
+ \mathbf{1}(d(r) < 0)\frac{d(r)}{ |\underline T(r)-T_0(r)| }.
\end{equation}
The {\em directional quantile} scaling \eqref{d_envdir} utilises as \eqref{d_env} the quantiles
$\underline T(r)$ and $\overline T(r)$, to weigh negative and
positive residuals, and, furthermore, as \eqref{d_env} it makes the variances
of residuals more uniform for the different distances $r$. The following toy example
demonstrates that severe asymmetry should be removed, if there is no
a priori interesting direction of deviation from $T_0(r)$.

{\em Toy example 2.} The random variables $X_i$ have asymmetric
distributions with distribution functions
\begin{equation}\label{Toyexample2_FXi}
F_{X_i}(x; \mu_i, \sigma_{ai}^2, \sigma_{bi}^2) = \frac{1}{2} F_a(x;
\mu_i, \sigma_{ai}^2) + \frac{1}{2} F_b(x; \mu_i, \sigma_{bi}^2),
\end{equation}
where $F_a$ is the distribution function of the truncated normal
distribution on $[\mu_i.\infty)$ and $F_b$ is the distribution
function for $(-\infty,\mu_i)$. The variances $\sigma_{ai}^2$ and
$\sigma_{bi}^2$ are assumed to be known. If
$\sigma_{ai}>\sigma_{bi}$ the fatter tail of the distribution of
$X_i$ lies right from $\mu_i$.

The null and alternative hypotheses are as in Toy example 1. Now the
power of the tests based on $U_{\infty} = \max_{i} |X_i|$ and
$U_{\infty,w} = \max_{i} \left(\mathbf{1}\left(X_i\geq 0\right)
|X_i|/a_{1i} + \mathbf{1}\left(X_i<0\right) |X_i|/a_{2i}\right)$
are compared. Again this can be carried out analytically, see the
Appendix.

\begin{figure}[h!]
\centering
\includegraphics{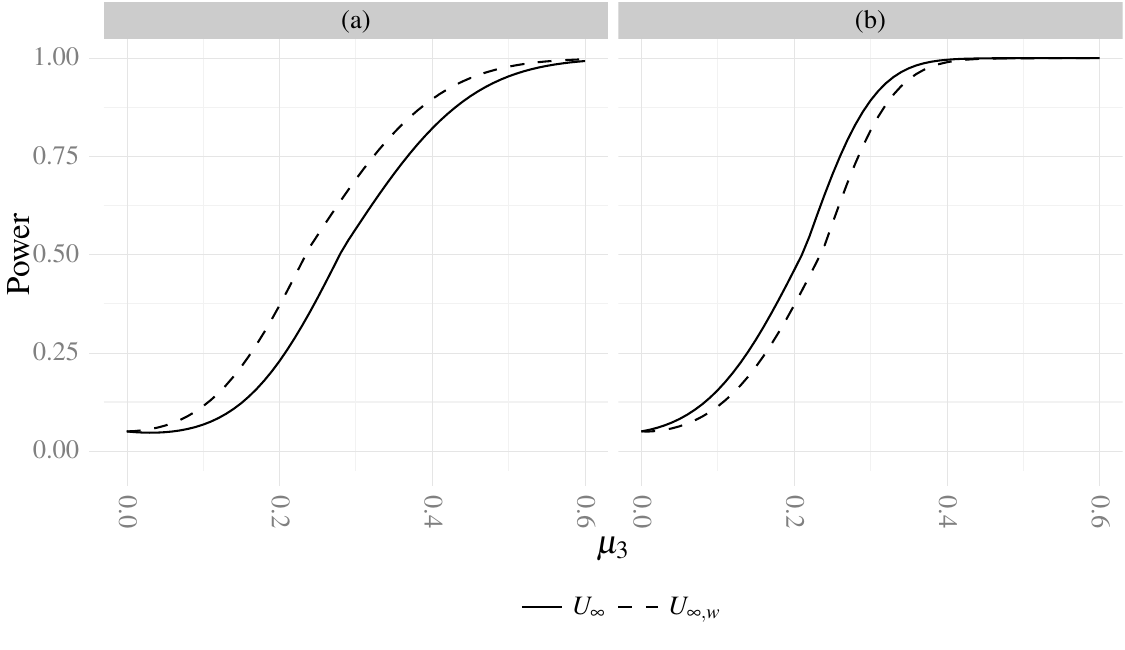}
\caption{Power curves as a function of $\mu_3$ for the alternative model $X_i\sim$\eqref{Toyexample2_FXi}
with $\mu_1=\mu_2=0$ and $\mu_3>0$ against the null hypothesis that $\mu_i = 0$ for all $i=1,2,3$.
The standard deviations in \eqref{Toyexample2_FXi} are $\sigma_{ai}=0.1$ and $\sigma_{bi}=c \sigma_{ai}$ for all $i=1,2,3$ with (a) $c=1.3$, (b) $c=0.7$. The lines represent the tests based on the deviation measures $U_{\infty}$ (unscaled test) and $U_{\infty,w}$ (scaled test). See the text for details.}\label{Fig_toyexample2}
\end{figure}

If $\sigma_{ai}<\sigma_{bi}$, the unscaled test has difficulties to
observe the positive deviation of $\mu_3$ from zero and has therefore smaller
power than the scaled test, see Figure \ref{Fig_toyexample2} (a).
On the other hand, for $\sigma_{ai}>\sigma_{bi}$ the scaling \eqref{d_envdir} is
counter productive, see Figure \ref{Fig_toyexample2} (b).

\section{Simulation study design}\label{sec:simulation_study_setup}

In order to be concrete, we now consider an important particular
case: testing of the random labelling hypothesis for marked point
patterns with non-negative real-valued marks. The assumption is
that the marked pattern can be thought to be generated in two steps:
first generating the points and then labelling the points
independently with marks following some mark distribution.
The corresponding point process
can be stationary, non-stationary or finite; the test is
carried out conditionally to the (unmarked) point pattern and, thus,
all these cases are treated identically.

As test functions $T(r)$ we use functions which are closely related
to the mark-weighted $K$-functions as defined for stationary marked
point processes in \citet{PenttinenStoyan1989} and
\citet{IllianEtal2008}. These functions are natural generalisations
of Ripley's $K$-function. We explain them first for the stationary
case.

\subsection{Mark-weighted $K$-functions}\label{sec:simstudy_summaryfunctions}

It is well known that Ripley's $K$-function can be explained as
follows: $\lambda K(r)$ is the mean number of other points within
distance $r$ from a typical point of the process, i.e.
$\lambda K(r)=\mathbb{E}_o\left( \sum_{x\in N} \mathbf{1}_{b(o,r)}(x) \right)$,
where the expectation is with respect to the Palm distribution and
$b(o,r)$ denotes the disc with radius $r$ centred at $o$
\citep{Ripley1976, Ripley1977}.

The mark-weighted $K$-function \citep{PenttinenStoyan1989,
IllianEtal2008}, $K_f(r)$, has a similar form as $K(r)$, but it also
takes the marks into account through a mark test function
$f(m_1,m_2): \mathbb{R}^2\rightarrow\mathbb{R_{+}}$. It is
\begin{equation*}
\lambda K_f(r)=\left.\mathbb{E}_o\left( \sum_{[x;\,m(x)]\in N_m} f(m(o),m(x))
\mathbf{1}_{b(o,r)}(x) \right) \right/c_f,
\end{equation*}
where \begin{equation}\label{cf}
 c_f = \int_0^{\infty} \int_0^{\infty} f(m_1,m_2) F_M(\text{d}m_1) F_M(\text{d}m_2),
\end{equation}
is a normalising factor, which depends on the test function $f$. In
\eqref{cf}, $F_M(m)$ is the mark distribution function.
In the case of independent marking it holds
\begin{equation}\label{=}
 K_f(r) = K(r) , \quad r\geq 0,
\end{equation}
see \citet{PenttinenStoyan1989}.

There are various possibilities for choosing the mark test function
$f$ \citep[see e.g.][]{Schlather2001, IllianEtal2008}. Table
\ref{Table:summaries} lists the $K_f$ functions that are used in
this paper, together with the corresponding $f$ and $c_f$.

The mark test functions in Table \ref{Table:summaries} look at mark behaviour from different
points of view. The function $K_{m.}$ explores the mark value given
that there is a further point within distance $r$. This function
may be used for detecting dependencies between marks and points
\citep{Schlatheretal2004, Guan2005}.
The function $K_{mm}(r)$ is based on products of marks of point
pairs of distance $r$. It is useful e.g.\ in situations where the
marks have the tendency to be smaller than the mean mark for points
close together. In contrast, $K_{\gamma}(r)$ is based on mark
differences and helps to detect situations where the marks of points
close together tend to be similar. Thus the choice of
$K_f$-functions for the test should be determined by knowledge of the patten.

\begin{table}[h!]
\begin{center}
\begin{tabular}{lll}
  Summary              & $f(m_1,m_2)$             &  $c_f$        \\\toprule
  $K_{m.}(r)$          & $m_1$                    &  $\mu_m$ \\
  $K_{mm}(r)$          & $m_1 m_2$                &  $\mu_m^2$ \\
  $K_{\gamma}(r)$      & $\frac{1}{2}(m_1 - m_2)^2$      &  $\sigma_m^2$ \\
\bottomrule
\end{tabular}
\caption{Different $K_f$-functions, their mark test functions
$f(m_1,m_2)$ and normalising factors $c_f$.}\label{Table:summaries}
\end{center}
\end{table}

The estimation of the $K_f$-functions in
the stationary case is carried out similarly as that of the
$K$-function. Let $n$ be the number of points of the analysed marked
point pattern observed in the window $W$ with area $|W|$. Then the
$K_f$-function can be estimated as
\begin{equation}\label{estKt}
\hat{K}_f(r) = \frac{1}{|W|\widehat{\lambda^2} \hat{c}_f} \sum_{k=1}^n\sum_{l=1, l\neq k}^n
f\left(m(x_k),m(x_l)\right)\mathbf{1}\left(||\,x_k-x_l||\leq
r\right) e(x_k,x_l),
\end{equation}
where $e(x_k,x_l)$ is an edge-correction factor and
\begin{equation}\label{cf_est}
\widehat{\lambda^2} = n(n-1) / |W|^2  \quad \text{and}\quad
\hat{c}_f = \sum_{i=1}^n\sum_{j=1, j\neq i}^n
f(m(x_i),m(x_j)) / n(n-1)
\end{equation}
are estimators of $\lambda^2$ and $c_f$, see \citet{IllianEtal2008}.
For the translational edge-correction, the edge-correction factor is
$e(x_k,x_l) = |W|/|W_{x_k}\cap W_{x_l}|$,
where $|W_{x_k}\cap W_{x_l}|$ is the area of the intersection of
$W_{x_k}$ and $W_{x_l}$, and $W_x$ is the translated window
$W_x=\{s+x,\, s\in W\}$ \citep[see][p.\,353]{IllianEtal2008}. For
the case of ``no edge-correction'', it is simply $e(x_k,x_l) \equiv
1$.

{\em Remark}. The estimator $\hat{c}_f$ in \eqref{cf_est}
differs in one point from the estimator
given in \citet[p.\ 353]{IllianEtal2008}:
the estimator \eqref{cf_est} is adapted to a marking in $W$ where
the marks of the $n$ points in $W$ are given to the points by random
permutation. This leads to the simple equation \eqref{randomlabelling_T0} below.
Note that in neither case $\hat{K}_f(r)$ is an
unbiased estimator of $K_f(r)$, because of the division by
$\hat{c}_f$ and $\widehat{\lambda^2}$.

\subsection{The test procedure}\label{sec:null_model}

The random labelling hypothesis is tested as follows.
Suppose a marked point pattern $\{[x_1, m(x_1)],\ldots, [x_n, m(x_n)]\}$ of
$n$ points is observed in the window $W$.
For this pattern the function $\hat{K}_f(r)$ is determined by \eqref{estKt}.
The result $\hat{K}_{f,1}(r)$ is compared with further functions
$\hat{K}_{f,i}(r)$ determined for $s$ simulated marked point patterns.
For all simulations the points are fixed, while the original marks $m(x_i)$
are randomly permuted. Thus $\hat{K}_f(r)$ plays the role of $T(r)$.

The expectation of $\hat{K}_f(r)$ under the null model of random
permutation of the marks is simply $\hat{K}(r)$, since
even $\hat{c}_f$ is fixed in \eqref{estKt} and only the $m(x_k)$ and $m(x_l)$ are variable.
This means that
\begin{equation}\label{randomlabelling_T0}
 T_0(r) = \hat{K}(r),
\end{equation}
where $\hat{K}(r)$ is obtained by \eqref{estKt} with $f\equiv 1$.
Note that this equation holds true for all forms of estimators of $K$ and $K_f$
as long as the same form of edge-correction,
e.g.\ translational, Ripley's or ``no correction'',
is used for $\hat{K}$ and $\hat{K}_f$.
Of course, the numerical values differ.
Note that all is conditional on the fixed points
in the window $W$, which implies that the test can be used also for
non-stationary or finite point processes.

To conduct the test, the residuals and global deviation measures are then
determined. In the following simulation study the effects of transformation,
scaling etc. are investigated.

\subsection{Alternative models}\label{sec:models}

For our power comparison we use four marked point process models
with different forms of mark correlations.
The first process is finite while the other three processes are stationary.

\subsubsection{Sequential neighbour-interaction marked point process}\label{sec:SeqNIMPP}

The {\it sequential neighbour-interaction marked point process}
(SeqNIMPP) can be thought of as a generalisation of Diggle's simple
sequential inhibition model known also as the random sequential
absorption (RSA) process \citep[see e.g.][]{PenroseShcherbakov2009}.
It is also related to the multivariate point process with
hierarchical interactions studied in \citet{GrabarnikSarkka2009},
who applied their model to an analysis of the spatial structure of a
forest stand. The SeqNIMPP model is a finite point process in $W$.

A realization in a bounded window $W$ can be constructed
as follows: Denote the marked point with location
$x$ and mark $m$ as $y=[x,m]$. Assuming that the marks of the marked
points $y_1, y_2, \ldots y_n$ are known, the corresponding locations
are allocated sequentially in $W$. The location $x_1$ of the first
point $y_1$ has uniform distribution over $W$, while the
location of the $k$th point follows the conditional density
$$
f_k( [x_k,m_k]\; |\; y_{k-1}, \ldots, y_1)=\frac{\exp\left\{ -
U_k([x_k,m_k]; \; y_{\leq k-1})\right\}} {\int_W \exp\left\{ -
U_k([x,m_k]; \;y_{\leq k-1})\right\} \text{d}x},
$$
where
$U_k(y_k; y_{\leq k-1})= \sum_{i=1}^{k-1} I(y_i, y_k)$.
The quantity $U_k$ can be
interpreted as the impact of the previously allocated points
$y_{\leq k-1}=(y_1, \ldots, y_{k-1})$ on the point $y_k=[x_k,m_k]$,
where $I(y_i, y_k)$ models the influence of the earlier point $y_i$
on the point $y_k$.

We assume that the marks stem from a Gaussian distribution with mean $\mu$ and
variance $\sigma^2$, truncated to avoid negative marks, and
use the influence function
\begin{equation*}
I(z,y)= \theta {\bf 1}\left( \|z-y\| < R\, m(z) / \mu\right) \frac{m(z)m(y)}{(\mu/R)^2 \|z-y\|}.
\end{equation*}
Here $m(z)$ is the mark of point $z$, $\theta$ is a parameter
controlling the strength of influence and $R$ is an interaction
radius.
If $\theta$ is positive, then the
points tend to avoid positions in the neighbourhood of previously
allocated points, the more the larger the marks of the previously
allocated points are. Further, the mark of the point itself
matters; the larger it is, the less likely it is
for it to appear close to a previously allocated point.
If $\theta$ is negative, on the other hand, then points
attract new points, and the degree of attraction depends on the marks of the
new and the previously allocated points.
The case $\theta=0$ corresponds to ``no
interaction between points'', i.e. independent marks.

\subsubsection{Exponential intensity-marked Cox process}\label{sec:ExpCP}

While the SeqNIMPP model creates regular patterns with inhibition
between points (for $\theta>0$), the stationary \emph{exponential intensity-marked
Cox process} (ExpCP) introduced by \citet{MyllymakiPenttinen2009}
forms clustered patterns of points. The points stem from a log
Gaussian Cox process (LGCP) $N=\{x_i\}$ with random intensity
$\Lambda(s) = \exp(Z(s))$, where $\{Z(s)\}$
is a stationary Gaussian random field. It has mean
$\mu_Z$ and the special covariance function $C_Z(r) =
\exp(-r/\phi_Z)$, where $\phi_Z$ is a range parameter. Conditional
on the random intensity the marks are distributed as
\begin{equation}\label{ExpIMCP_neg_marks}
 m(x_i) | \Lambda(x_i) \sim \text{Exp}\Big(1 / (a+ b/\Lambda(x_i))\Big)
\quad \text{(ExpNIMCP)}
\end{equation}
with expectation $\mathbb{E}(m(x_i) | \Lambda(x_i)) = a+ b/\Lambda(x_i)$,
or by
\begin{equation}\label{ExpIMCP_pos_marks}
 m(x_i) | \Lambda(x_i) \sim \text{Exp}\Big(1 / (a+ b \Lambda(x_i))\Big)
\quad \text{(ExpPIMCP)}
\end{equation}
with expectation $\mathbb{E}(m(x_i) | \Lambda(x_i)) = a+ b \Lambda(x_i)$,
where $a$ and $b$ are positive model parameters.
In these models both the mean and variance of marks depend on
the local point density: In the ExpNIMCP model, the marks tend to be small
and less variable in areas with high intensity, while in low
intensity areas both small and large marks can occur. In the ExpPIMCP
model, on the other hand, mean and variance of marks tend to be large
in high intensity areas. For more details see
\citet{MyllymakiPenttinen2009} and \citet{Myllymaki2009}, where the
model \eqref{ExpIMCP_neg_marks} was applied to model structure of
rainforest data.

\subsubsection{Gaussian noise intensity-marked Cox process}\label{sec:GNIMCP}

The \emph{Gaussian noise intensity-marked Cox process} (GNIMCP) is
similar to the ExpIMCP model and the models considered in
\citet{HoStoyan2008}. The points stem from the same LGCP model with
the intensity $\Lambda(s)=\exp(Z(s))$ as the points in the ExpIMCP model,
and the marks are
\begin{equation*}
 m(x_i) = a \cdot \exp\left\{ b \left(\frac{Z^*(x_i)- \mu_Z}{1 + \sigma_{\epsilon}} \right) \right\},
\end{equation*}
where $a$ and $b$ are real mark-scale parameters and $\mu_Z$ is the mean of $Z$.
The strength of dependence between marks and points is controlled through $Z^*$:
$Z^*(x_i) = Z(x_i) + \epsilon(x_i)$,
where $\epsilon(x_i) \sim N(0, \sigma_{\epsilon}^2)$ are i.i.d.
The larger the variance $\sigma_{\epsilon}^2$ is,
the noisier the values of $Z^*$ and marks are.

If the parameter $b$ is positive (negative), then the marks in
high intensity areas tend to be larger (smaller) than in low
intensity areas. The distribution of $m(x_i)$ given $Z(x_i)$ is
lognormal.

\subsubsection{A random field model: Gaussian noise Cox process}\label{sec:GNCP}

Again, the points come from the LGCP model. The \emph{Gaussian noise
Cox process} (GNCP) is a particular case of the random field model,
see \citet{IllianEtal2008}, since the marks are generated as
follows: To mimic the GNIMCP model, let $\{Z_M(s)\}$ be an
independent Gaussian random field with mean $\mu_Z$ and covariance
function $C_Z(r)=\exp(-r/\phi_Z)$, and define
$Z_M^*(s) = Z_M(s) + \epsilon(s)$
and
\begin{equation*}
U(s) = a\cdot \exp\left\{ b \left(\frac{Z_M^*(s)- \mu_Z}{1 + \sigma_{\epsilon}} \right) \right\},
\end{equation*}
where $a$ and $b$ are model parameters and $\epsilon(s)\sim N(0, \sigma_{\epsilon}^2)$.
Then the marks are simply
$m(x_i) = U(x_i)$.

In this model, there is no relationship between local point density and marks, but
points close together tend to have similar marks.
This similarity of marks decreases when the variance $\sigma_{\epsilon}^2$ increases.

\section{Power comparison by the simulation study}\label{sec:simstudy}

In order to observe how the ingredients of the random labelling test
affect its power, we tested the random labelling hypothesis against
the four alternative marked point process models in Section
\ref{sec:models} with various parameter combinations summarised in
Table \ref{Table:simexp_models}. In each model, one of the
parameters (called the ``changing parameter'') controlling
the strength of spatial correlations
has been selected and is systematically varied, while the other, fixed
parameters have been chosen such that the most prominent deviation of
$T(r)$ from $T_0(r)$ occurs approximately at the distance $r=6$
(determined by simulation).

\begin{table}[h!]
\begin{center}
\begin{tabular}{lll}
Model           & Fixed mark parameters            &  Changing parameter         \\\toprule
SeqNIMPP        & $\mu=24$, $\sigma^2=9$                      &  $\theta=0,0.02,0.04,\dots,0.20$ \\
ExpNIMCP \eqref{ExpIMCP_neg_marks}  & $b=1$     & $a = 0, 20, 40, \dots, 200$ \\
ExpPIMCP \eqref{ExpIMCP_pos_marks}  & $b=6600$  & $a = 0, 250, 500, \dots, 2500$ \\
GNIMCP, $b<0$   & $a=24$, $b=-0.12$     & $\sigma_{\epsilon} = 0,0.5,1.0,\dots, 6.0$ \\
GNIMCP, $b>0$   & $a=24$, $b=0.12$      & $\sigma_{\epsilon} = 0,0.5,1.0,\dots,6.0$ \\
GNCP            & $a=24$, $b=-0.12$     & $\sigma_{\epsilon} = 0,0.25,0.5,\dots,3.5$ \\
\bottomrule
\end{tabular}
\caption{Parameter values of the SeqNIMPP (see Section \ref{sec:SeqNIMPP}), ExpIMCP (Section \ref{sec:ExpCP}), GNIMCP (Section \ref{sec:GNIMCP}) and GNCP (Section \ref{sec:GNCP}) models used in the simulation experiment. As the mean and range of correlation parameters of the LGCP models we used $\mu_Z=-4.4$ and $\phi_Z=4$, respectively.
}\label{Table:simexp_models}
\end{center}
\end{table}

For each model we made $N=1000$ simulations with $n=200$ points in a window of size
$[0,100]\times[0,100]$.
For each simulated marked point pattern, we
performed tests based on $s=999$ random permutations of marks. The
tests were made with
\begin{enumerate}
\renewcommand{\theenumi}{\roman{enumi}}
\renewcommand{\labelenumi}{(\theenumi)}
 \item $T(r)=\hat{K}_f(r)$ (\eqref{estKt} with the translational edge correction)
with the mark test functions $f$ given in Table \ref{Table:summaries},
 \item the transformation $h(\cdot)=\sqrt{\cdot/\pi}$,
 \item raw \eqref{rawresidual}, studentised \eqref{d_var}, quantile \eqref{d_env}
and directional quantile \eqref{d_envdir} residuals,
 \item deviation measures \eqref{Dinfty} and \eqref{DL2}, and
 \item three intervals $I$ of $r$-values: $I_1=[4,8]$, $I_2=[3,15]$ and $I_3 = [0,25]$.
\end{enumerate}
All alternatives in (iii)-(v) were considered both for
the test function $T(r) = \hat{K}_f(r)$,
and its transformation $\hat{L}_f(r) = \sqrt{\hat{K}_f(r)/\pi}$.
The three intervals $I$ were chosen to separate effects which could
influence the power in the comparative study.

We argue that a fair comparison between different forms
of the deviation test can be done only if the distributions of residuals
are approximately uniform.
Therefore,
we first compare the different mark test functions based on the
shortest interval $I_1$, in order to eliminate effects of the other factors,
and choose the most powerful mark test function for each model,
which will be used in other comparisons.
We then use the widest interval $I_3$ for comparing (ii)-(iii),
and also (iv) after scalings,
while $I_2$ is used to study the effects of the width of $I$.

The number of simulated patterns for which the null hypothesis is
rejected among the $N$ simulations gives an estimate for the
power. The results below are for the significance level 0.05.

\subsection{Comparison of mark test functions}\label{sec:simstudy_marktestfs}

\begin{figure}[h!]
\centering
\includegraphics{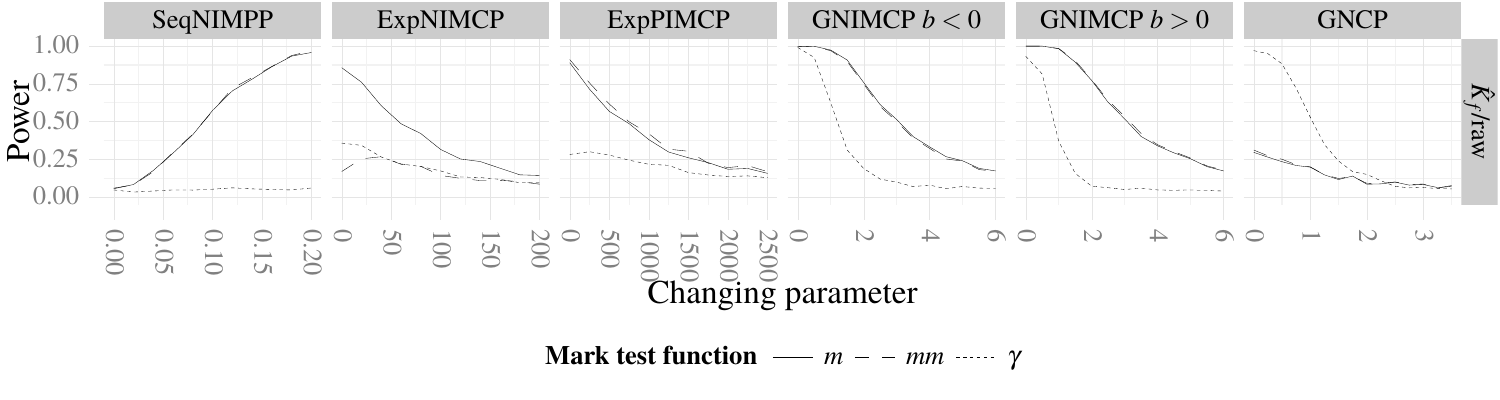}
\caption{Power comparison of the $\hat{K}_f$ functions with different mark test functions $f(m_1,m_2)$ for the models of Table \ref{Table:simexp_models} (top panel) using the test based on the supremum deviation measure \eqref{Dinfty} applied to the raw residuals of $\hat{K}_f(r)$ on $I_1=[4,8]$.}\label{Fig_simexp_marktestfs}
\end{figure}

As said above, to compare different mark test functions,
the attention is restricted to the narrow interval $I_1=[4,8]$.
Since the contributions of raw residuals of $\hat{K}_f(r)$
for different $r$ are approximately equal on $I_1$,
the test based on these residuals can be used without any transformations or scalings.
On this narrow interval, also differences between
the deviation measures \eqref{Dinfty} and \eqref{DL2}
are negligible and, thus, Figure \ref{Fig_simexp_marktestfs}
shows the results only for \eqref{Dinfty}.
The power for the various summary functions $\hat{K}_f$
(or mark test functions $f$) depends on the alternative model:

\begin{enumerate}

\item The function $\hat{K}_{m.}$ leads to powers at least as high as for
any other function for the SeqNIMPP, ExpNIMCP and GNIMCP models,
while, for the random field model GNCP, its power is low.
For the ExpPIMCP model, the power related to $\hat{K}_{m.}$ is slightly
lower than that for $\hat{K}_{mm}$.

\item The function $\hat{K}_{mm}$ is approximately as powerful as
$\hat{K}_{m.}$ for the SeqNIMPP, ExpPIMCP and GNIMCP models, where the marks
tend to be either smaller or larger (depending on model parameters)
than the mean mark for points close together.
On the other hand,
in the ExpNIMCP model, the marks at points close together
are not clearly smaller than the mean mark and, therefore,
$\hat{K}_{mm}$ does not have high power.
For the GNCP model, the power is similarly low.

\item The function $\hat{K}_{\gamma}$ leads to a powerful
test for the GNCP model, because of its sensibility to similar marks
at short distances.
Note that the marks of points close together also tend to be
similar also in the GNIMCP models, but for these models
$\hat{K}_{m.}$ and $\hat{K}_{mm}$
lead to much more powerful test than $\hat{K}_{\gamma}$ for the reasons
explained above.

\end{enumerate}

We conclude that indeed the power of deviation test depends on the
choice of the summary function, which should be adapted to the
alternative model.

\subsection{Comparison of transformations and scalings}\label{sec:simstudy_scalings}

\subsubsection{Transformation}

\begin{figure}[h!]
\centering
\includegraphics{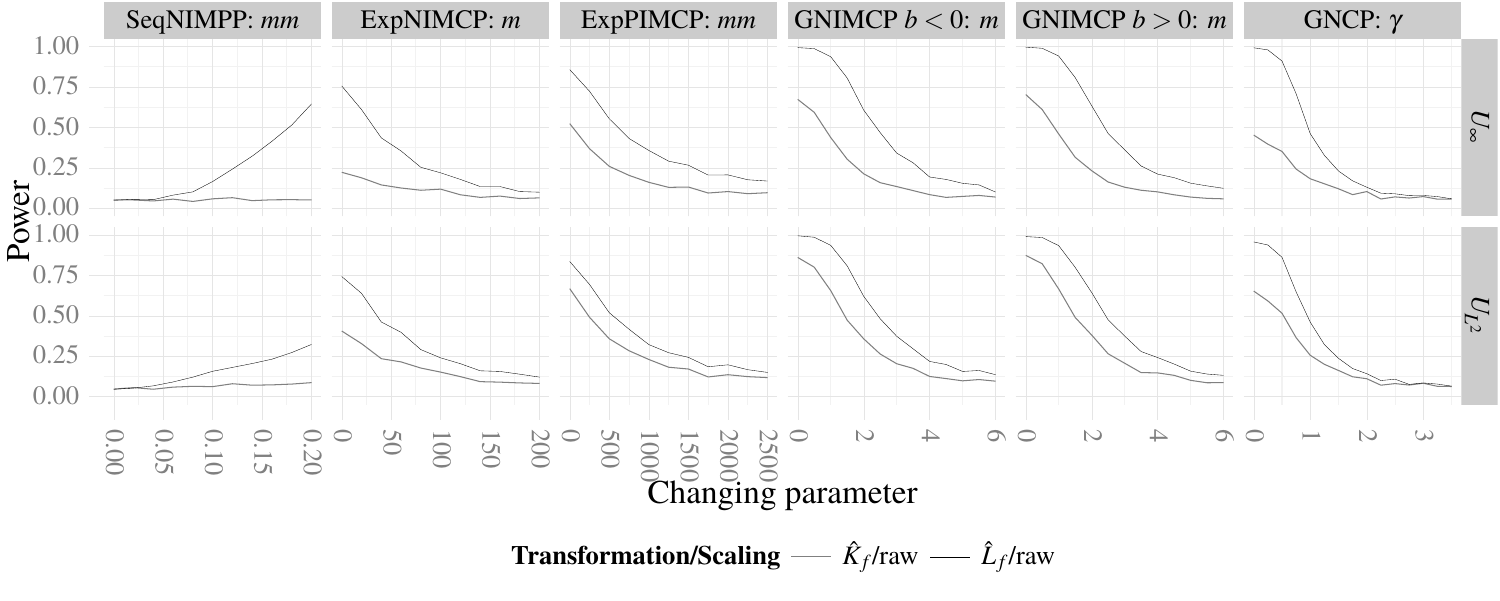}
\caption{Power comparison of square root transformation for the models of Table \ref{Table:simexp_models} with the most powerful mark test functions $f$ (top panel) using the deviation measures \eqref{Dinfty} and \eqref{DL2} (right panel) applied to the raw residuals of $\hat{K}_f(r)$ and $\hat{L}_f(r)$ on $I_3=[0,25]$.}\label{Fig_simexp_scalings_KvsL}
\end{figure}

The square root transformation increases greatly the power
for all the alternative models,
as it is seen from the power curves of the tests based on raw residuals of
$\hat{K}_f(r)$ and $\hat{L}_f(r)$ in Figure \ref{Fig_simexp_scalings_KvsL}.
The reason for the increase in power is that the residuals of $\hat{L}_f(r)$
are more uniform than those of $\hat{K}_f(r)$.

\subsubsection{Scaling}

\begin{figure}[h!]
\centering
\includegraphics{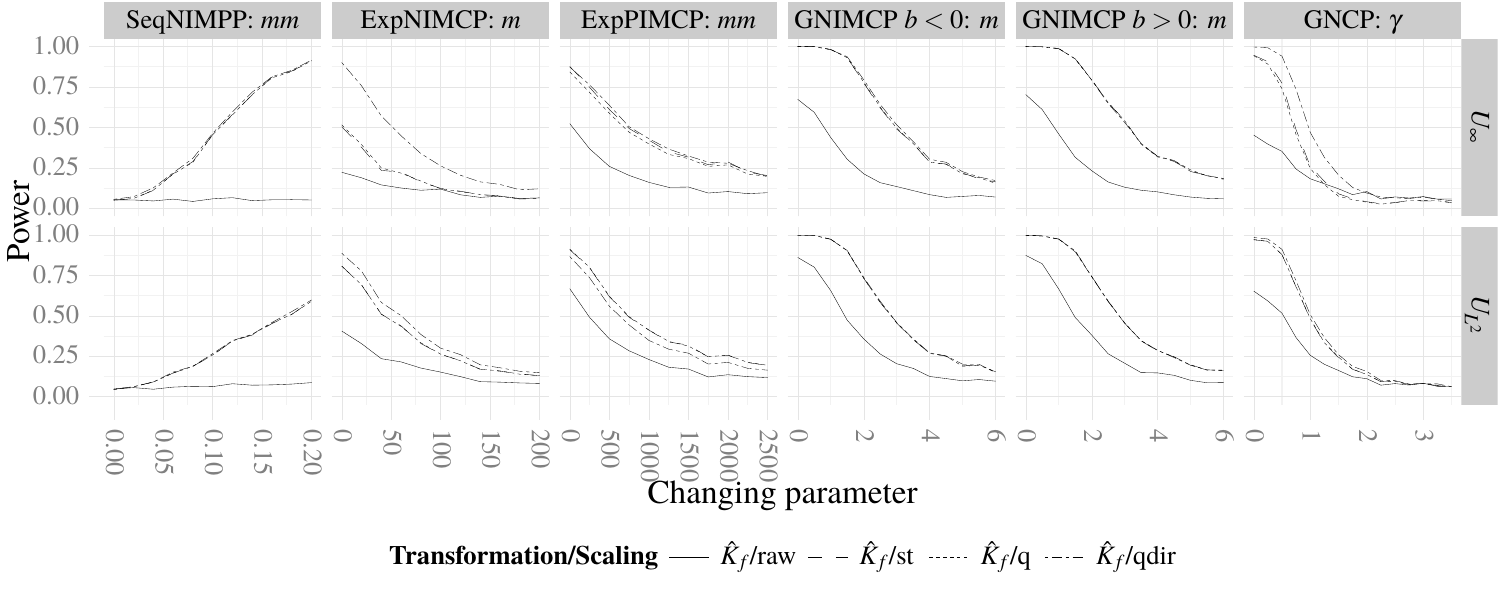}
\caption{Power comparison of scalings (raw; studentised 'st', quantile 'q' and directional quantile 'qdir') for the models of Table \ref{Table:simexp_models} with the most powerful mark test functions $f$ (top panel) using the deviation measures \eqref{Dinfty} and \eqref{DL2} (right panel) applied to the different residuals of $\hat{K}_f(r)$ on $I_3=[0,25]$.}\label{Fig_simexp_scalings_Kbased}
\end{figure}

\begin{figure}[h!]
\centering
\includegraphics{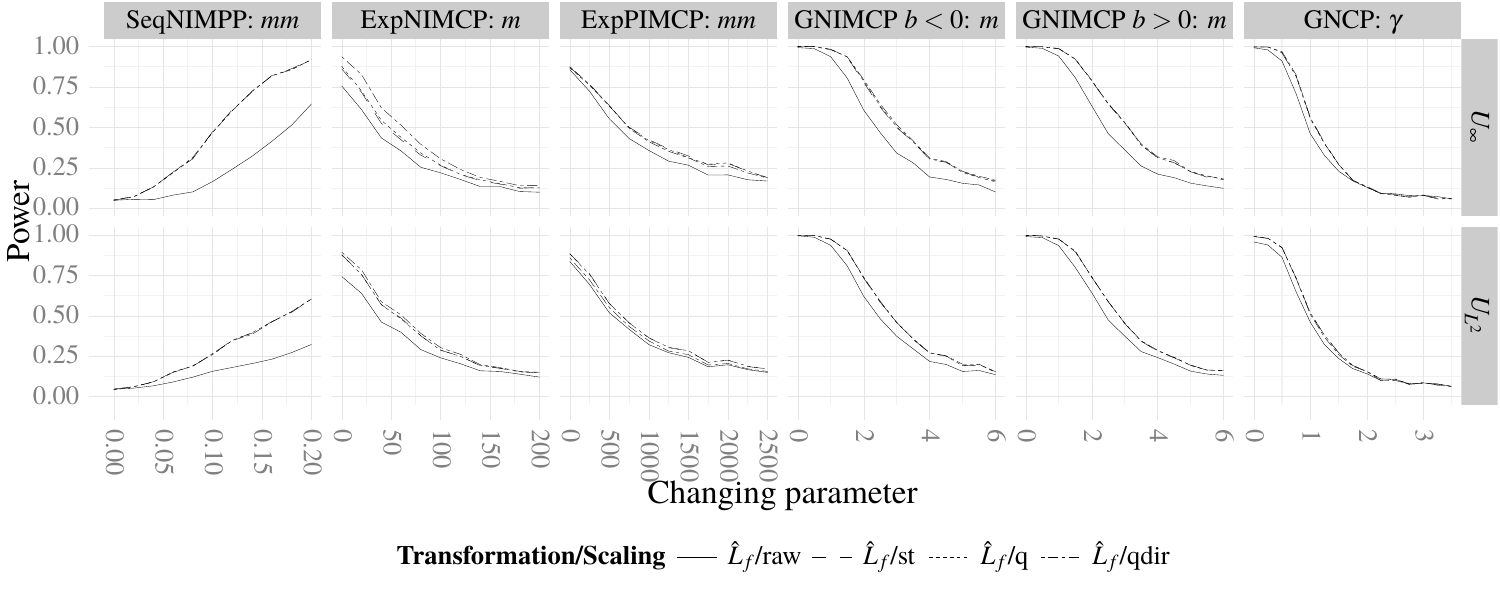}
\caption{Power comparison of scalings (raw; studentised 'st', quantile 'q' and directional quantile 'qdir') for the models of Table \ref{Table:simexp_models} with the most powerful mark test functions $f$ (top panel) using the deviation measures \eqref{Dinfty} and \eqref{DL2} (right panel) applied to the different residuals of $\hat{L}_f(r)$ on $I_3=[0,25]$.}\label{Fig_simexp_scalings_Lbased}
\end{figure}

Figures \ref{Fig_simexp_scalings_Kbased} and
\ref{Fig_simexp_scalings_Lbased} show power curves for
the deviation measures \eqref{Dinfty} and \eqref{DL2} applied to the raw \eqref{rawresidual},
studentised \eqref{d_var}, quantile \eqref{d_env} and directional quantile \eqref{d_envdir}
residuals of the test function $T(r)=\hat{K}_f(r)$ and its transformation $\hat{L}_f(r)$, respectively.
We found the following:
\begin{enumerate}
 \item Always the power of the tests based on raw residuals is lowest.

 \item The results with the studentised \eqref{d_var} and
quantile \eqref{d_env} scalings are very similar.

  \item For the ExpNIMCP and GNCP models (for the latter only for
$\hat{K}_f(r)$), the directional scaling \eqref{d_envdir} improves further
the power, while for the ExpPIMCP the scaling
\eqref{d_envdir} is counter productive.
The problem of asymmetry plays a role for these models:
The ExpIMCP models have an asymmetric mark distribution, and
we found that also the empirical distributions of
residuals are asymmetric. Moreover, we found that the
residual distributions are more asymmetric for $\hat{K}_{\gamma}$
than for $\hat{K}_{m.}$ and $\hat{K}_{mm}$, which
explains the result for the GNCP model.
\end{enumerate}
Note that the improvements of power by
scalings are smaller for $\hat{L}_f(r)$ than for $\hat{K}_f(r)$, since
the residuals of the transformed test function are more uniform
than the raw residuals.

\subsubsection{Combined transformation and scaling}

\begin{figure}[h!]
\centering
\includegraphics{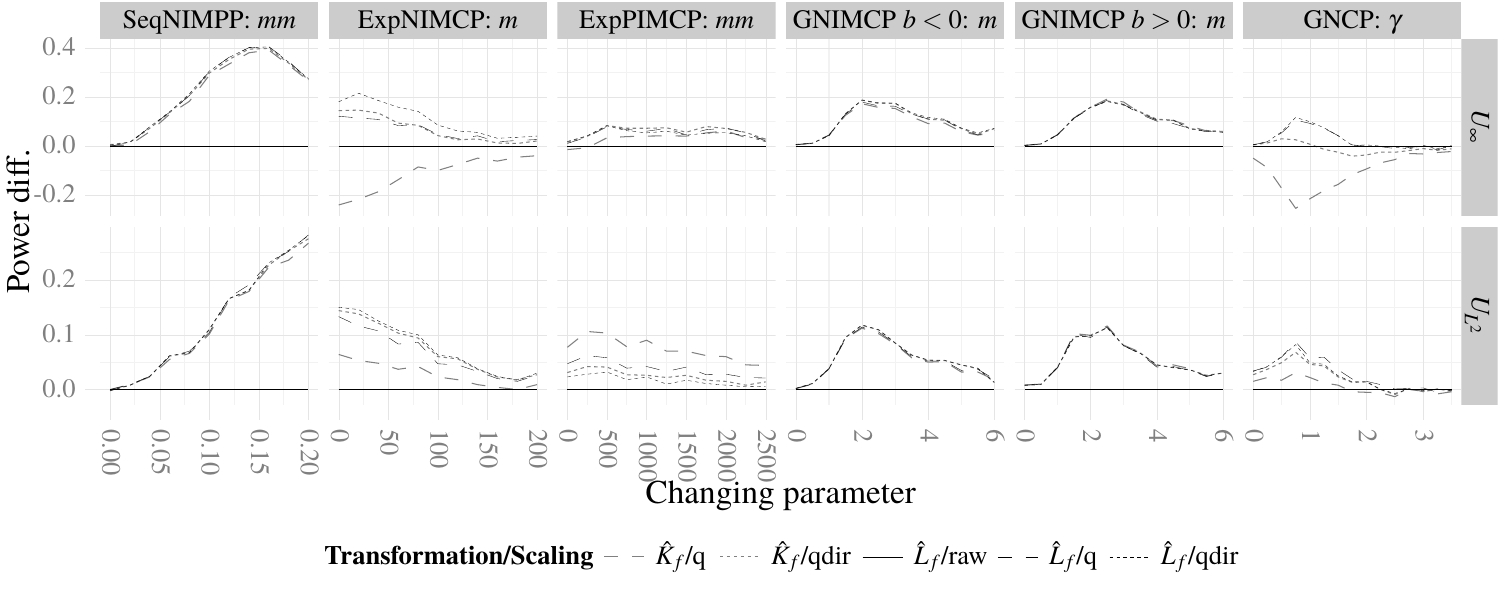}
\caption{Power comparison of the square root transformation and scalings (raw; studentised 'st', quantile 'q' and directional quantile 'qdir') for the models of Table \ref{Table:simexp_models} with the most powerful mark test functions (top panel) using the deviation measures \eqref{Dinfty} and \eqref{DL2} (right panel) on $I_3=[0,25]$. The curves show the power difference with respect to the power of the test based on raw residuals of $\hat{L}_f$ (represented by the zero line).}\label{Fig_simexp_scalings_vsL}
\end{figure}

Figure \ref{Fig_simexp_scalings_vsL} compares the tests based on
the scaled residuals of $\hat{K}_f$ to those of $\hat{L}_f$.
We observe the following:
\begin{enumerate}
\item For the SeqNIMPP and GNIMCP models, the differences between
the powers of the tests based on scaled residuals of $\hat{K}_f(r)$
or $\hat{L}_f(r)$ are very small, and these tests
have clearly higher power than the test based on the raw residuals of
the transformed test function $\hat{L}_f(r)$.

\item For the ExpNIMCP and GNCP models,
the power for the test based on the quantile residuals \eqref{d_env}
of $\hat{K}_f(r)$ is lower than for the test based on the corresponding
residuals of $\hat{L}_f(r)$ (and, for the measure \eqref{Dinfty},
even lower than for the test based on raw residuals of $\hat{L}_f$).
For the ExpPIMCP model, the opposite occurs for the measure \eqref{DL2}.

\item Similarly, applying first the square root transformation
and then the scaling \eqref{d_envdir} results in more powerful
tests than applying pure scaling \eqref{d_envdir} to $\hat{K}_f(r)$
for the ExpNIMCP and GNCP models, and the other way around for the
ExpPIMCP model.

\item For the GNCP model, the scaling \eqref{d_envdir} has no
advantage over the scaling \eqref{d_env} if the transformation is employed first.
\end{enumerate}

The result 1 above indicates that the transformation, prior to scaling, is unnecessary
for the SeqNIMPP and GNIMCP models, whereas the results 2-3 show that it is
useful in the case of the ExpNIMCP and GNCP models.
As pointed out already above, for the latter models the asymmetry plays a role.
Clearly the scalings \eqref{d_env} and \eqref{d_var} do not decrease
asymmetry in the distribution of the residuals, but, in this particular case,
it appears that the square root transformation does,
as does the scaling \eqref{d_envdir}.

Thus, a transformation can lead to a change in the form of the
distribution of residuals. In general, it can reduce as well as induce asymmetry.
However, typically it will not make the distribution
of residuals completely uniform for all distances on $I$ and
scalings lead to further improvements in power, see
Figures \ref{Fig_simexp_scalings_Lbased} and \ref{Fig_simexp_scalings_vsL}.
We conclude that a good strategy appears to be to use
a suitable transformation, if available, and then scaling to further reduce
inhomogeneity of residuals.

\subsection{Comparison of deviation measures}\label{sec:simstudy_devmeasures}

\begin{figure}[h!]
\centering
\includegraphics{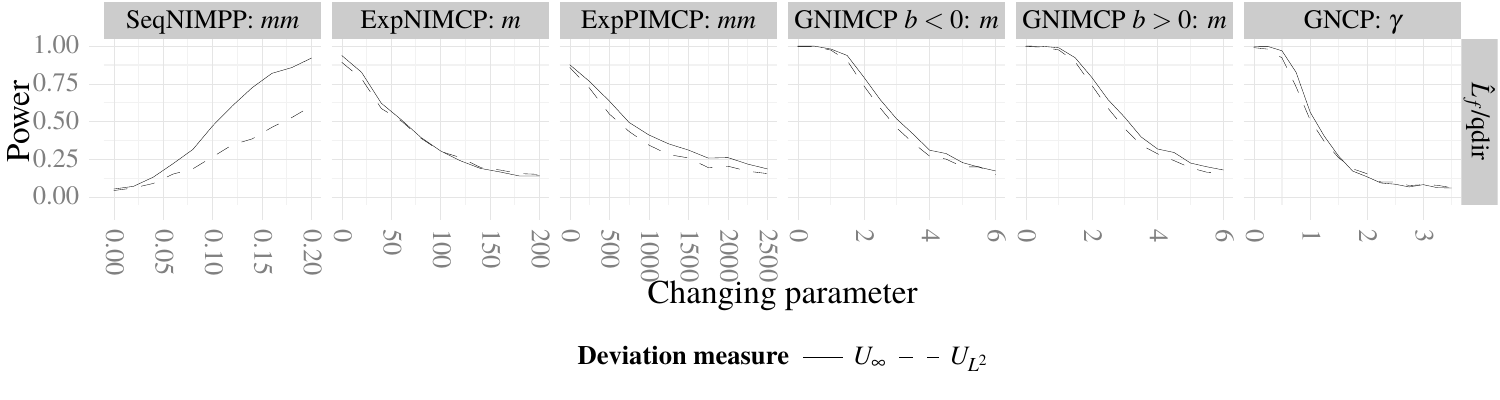}
\caption{Power comparison of the supremum $U_{\infty}$ \eqref{Dinfty} and integral $U_{L^2}$ \eqref{DL2} deviation measures applied to the residuals \eqref{d_envdir} of $\hat{L}_f$ on $I_3=[0,25]$ for the models of Table \ref{Table:simexp_models} with the most powerful mark test functions (top panel).}\label{Fig_simexp_devmeasures}
\end{figure}

Figure \ref{Fig_simexp_devmeasures} shows results with the supremum
\eqref{Dinfty} and integral \eqref{DL2} deviation measures applied
to the scaled residuals \eqref{d_envdir} of $\hat{L}_f$ on the range
of distances $I_3=[0,25]$.
For the SeqNIMPP model, for which the deviance from the null model is reasonably sharp,
the supremum measure leads to a clearly higher power than the integral measure.
For the other models, the measures \eqref{Dinfty} and \eqref{DL2}
lead approximately to the same power and, thus, do not play an important role.

\subsection{Range of distances and scaling}\label{sec:simstudy_rangeofdist}

\begin{figure}[h!]
\centering
\includegraphics{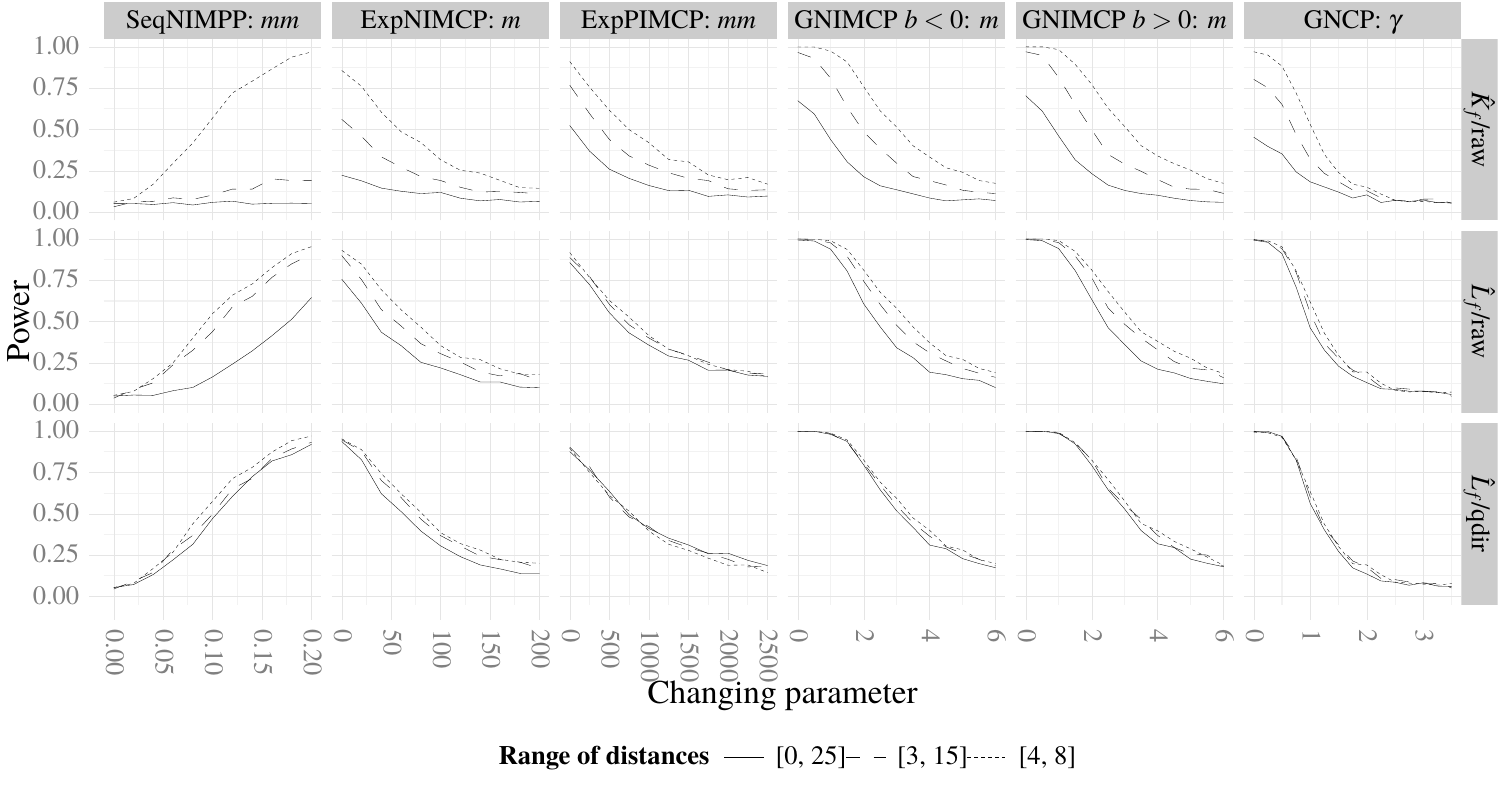}
\caption{Power comparison of the intervals of distances $I_3=[0,25]$, $I_2=[3,15]$ and $I_1=[4,8]$ for the models of Table \ref{Table:simexp_models} with the most powerful mark test functions (top panel) using the supremum deviation measure \eqref{Dinfty} applied to the raw residuals of $\hat{K}_f$ and $\hat{L}_f$ and scaled residuals \eqref{d_envdir} of $\hat{L}_f$ (right panel).}\label{Fig_simexp_rangeofdist}
\end{figure}

There is some balance between the length of the interval $I$ and
scaling. Figure \ref{Fig_simexp_rangeofdist} shows powers for tests
with different intervals $I$. The powers of the tests based on raw
residuals of $\hat{K}_f$ as well as for $\hat{L}_f$ are highly
increased from $I_3$ to $I_2$ and from $I_2$ to $I_1$. However, for
the tests based on scaled residuals we observe hardly any increase
in power from $I_3$ and $I_2$ to $I_1$.

In general, the shorter the interval $I$, the more powerful the test
can be as long as the interesting behavior occurs inside $I$.
However, as the results in Figure \ref{Fig_simexp_rangeofdist} show,
if appropriate transformations and scalings are used,
there is only a weak relationship between the power and the length of $I$.
However, for the SeqNIMPP model, the non-powerful deviation measure \eqref{DL2}
is still clearly more powerful on $I_1$ than on $I_2$ and $I_3$
(figure not shown).

At the end of this section we mention briefly the two further results.
The first concerns the problem of edge-correction.
As discussed already, the estimator \eqref{estKt}
stems from the stationary process context. In the
random labelling test there is no need of edge-correction since we
deal with a fixed point pattern in the bounded window $W$.
So we also made the tests on $I_3$
for the processes in Table \ref{Table:simexp_models}
without edge-correction
(i.e.\ $e(x_k, x_l)=1$) and observed that the power was a little
higher than for the estimator with edge-correction if no scalings
were used.

We also made a corresponding simulation experiment as reported above
based on $n=800$ points in a window of size $[0,200]\times[0,200]$.
The results with $n=800$ were analogous to those with $n=200$,
except the values of power were larger, showing the
consistency property empirically.

\section{Discussion and conclusions}\label{sec:discussion}

This paper considers the construction of deviation tests for point
processes in a general form.
The tests are applicable also for non-stationary and finite point processes,
since they are based on a test function and its expectation under the null hypothesis,
which can be estimated from simulations from the null model if it is not known analytically.
Thus, there is no need to assume stationarity and
to use unbiased estimators of summary functions in the test.
The paper demonstrates the tests for finite and stationary point processes in the case
of the random labelling hypothesis.

The main point in constructing a powerful deviation test is to choose
a suitable test function $T(r)$. For the random labelling test we used
mark-weighted $K$ functions, where the choice of the mark test function
$f$ is essential. Which mark test functions will lead to the most powerful test
depends on properties of the point pattern under analysis.
If a researcher has an alternative model in mind, this model may suggest
the mark test function to be used.

When the test function $T(r)$ is chosen, the next problem is
search for a suitable transformation. The simulation study shows that
transformations of summary functions can increase the power of
deviation tests, since they allow to make the distribution of residuals
more uniform over distances.
There are some classical transformations, and
perhaps new transformations can be developed.

Finally, scalings of residuals can further improve the power of tests based on transformed
summary functions, because they lead to approximately
even variances of residuals for different distances, unlike transformation only.
While transformations may reduce or induce asymmetry in the
distribution of residuals, scaling is a simple way to make distributions
of residuals more uniform.
As shown by the simulation study, scalings also act in some form of balance with the choice
of the length of the interval $I$, i.e.\ 
an appropriate scaling can make the choice of $I$ unimportant.

In some sense, the role of transformations and scalings is
similar to that of prior distributions in Bayesian statistics. If
there is not an a priori interesting distance $r$, then it makes
sense to give similar importance to all residuals in the chosen interval of distances $I$.
Thus, the residual distributions for different distances should be
made uniform, which, in the context of deviation tests,
can be done by means of transformations and scalings.
As the toy examples and the simulation study demonstrated,
such transformations and scalings typically, but not necessarily,
improve the power of the deviation test.

\section*{Acknowledgements}

M.\,M. has been financially supported by the Academy of Finland (project number 250860)
and P.\,G. by RFBR grant (project 12-04-01527).
The authors thank Tom\' a\v s Mrkvi\v cka (University of South Bohemia)
for his comments on an earlier version of the article.


\begin{thebibliography}{32}
\expandafter\ifx\csname natexlab\endcsname\relax\def\natexlab#1{#1}\fi

\bibitem[{Aitkin and Clayton(1980)}]{AitkinClayton1980}
Aitkin, M. and Clayton, D. (1980).
\newblock The fitting of exponential, Weibull and extreme value distributions
  to complex censored survival data using glim.
\newblock \emph{Applied Statistics} \textbf{29}, 156--163.

\bibitem[{Baddeley \emph{et~al.}(2000)Baddeley, Kerscher, Schladitz and
  Scott}]{BaddeleyEtal2000b}
Baddeley, A.~J., Kerscher, M., Schladitz, K. and Scott, B.~T. (2000).
\newblock Estimating the $J$ function without edge correction.
\newblock \emph{Stat. Neerl.} 
\textbf{54}, 315--328.

\bibitem[{Barnard(1963)}]{Barnard1963}
Barnard, G.~A. (1963).
\newblock Discussion of professor {Bartlett's} paper.
\newblock \emph{J. R. Stat. Soc. Ser. B Stat. Methodol.} 
\textbf{25}, 294.

\bibitem[{Besag and Diggle(1977)}]{BesagDiggle1977}
Besag, J. and Diggle, P.~J. (1977).
\newblock Simple Monte Carlo tests for spatial pattern.
\newblock \emph{J. R. Stat. Soc. Ser. C. Appl. Stat.} 
\textbf{26}, 327--333.

\bibitem[{Besag(1977)}]{Besag1977}
Besag, J.~E. (1977).
\newblock Comment on `{Modelling spatial patterns}' by {B. D. Ripley}.
\newblock \emph{J. R. Stat. Soc. Ser. B. Stat. Methodol.} 
\textbf{39}, 193--195.

\bibitem[{Bretz \emph{et~al.}(2010)Bretz, Hothorn and Westfall}]{BretzEtal2010}
Bretz, F., Hothorn, T. and Westfall, P. (2010).
\newblock \emph{Multiple comparisons using R}.
\newblock Chapman and {Hall/CRC}, 1st edn.

\bibitem[{Cressie(1993)}]{Cressie1993}
Cressie, N. A.~C. (1993).
\newblock \emph{Statistics for spatial data}, revised edn.
\newblock 
Wiley, New York.

\bibitem[{Diggle(1979)}]{Diggle1979}
Diggle, P.~J. (1979).
\newblock On parameter estimation and goodness-of-fit testing for spatial point
  patterns.
\newblock \emph{Biometrics} \textbf{35}, 87--101.

\bibitem[{Diggle(2003)}]{Diggle2003}
Diggle, P.~J. (2003).
\newblock \emph{Statistical analysis of spatial point patterns}, 2nd edn.
\newblock Arnold, London.

\bibitem[{Gignoux \emph{et~al.}(1999)Gignoux, Duby and Barot}]{Gignouxetal1999}
Gignoux, J., Duby, C. and Barot, S. (1999).
\newblock Comparing the performances of Diggle's tests of spatial randomness
  for small samples with and without edge-effect correction: Application to
  ecological data.
\newblock \emph{Biometrics} \textbf{55}, 156--164.

\bibitem[{Grabarnik and Chiu(2002)}]{GrabarnikChiu2002}
Grabarnik, P. and Chiu, S.~N. (2002).
\newblock Goodness-of-fit test for complete spatial randomness against mixtures
  of regular and clustered spatial point processes.
\newblock \emph{Biometrika} \textbf{89}, 411--421.

\bibitem[{Grabarnik \emph{et~al.}(2011)Grabarnik, Myllym{\"a}ki and
  Stoyan}]{GrabarnikEtal2011}
Grabarnik, P., Myllym{\"a}ki, M. and Stoyan, D. (2011).
\newblock Correct testing of mark independence for marked point patterns.
\newblock \emph{Ecological Modelling} \textbf{222}, 3888--3894.

\bibitem[{Grabarnik and S{\"a}rkk{\"a}(2009)}]{GrabarnikSarkka2009}
Grabarnik, P. and S{\"a}rkk{\"a}, A. (2009).
\newblock Modelling the spatial structure of forest stands by multivariate
  point processes with hierarchical interactions.
\newblock \emph{Ecological Modelling} \textbf{220}, 1232--1240.

\bibitem[{Guan(2005)}]{Guan2005}
Guan, Y. (2005).
\newblock Tests for independence between marks and points of a marked point
  process.
\newblock \emph{Biometrics} \textbf{62}, 126--134.

\bibitem[{Ho and Chiu(2006)}]{HoChiu2006}
Ho, L.~P. and Chiu, S.~N. (2006).
\newblock Testing the complete spatial randomness by {Diggle's} test without an
  arbitrary upper limit.
\newblock \emph{J. Stat. Comput. Simul.} 
\textbf{76}, 585--591.

\bibitem[{Ho and Chiu(2009)}]{HoChiu2009}
Ho, L.~P. and Chiu, S.~N. (2009).
\newblock Using weight functions in spatial point pattern analysis with
  application to plant ecology data.
\newblock \emph{Comm. Statist. Simulation Comput.} 
\textbf{38}, 269--287.

\bibitem[{Ho and Stoyan(2008)}]{HoStoyan2008}
Ho, L.~P. and Stoyan, D. (2008).
\newblock Modelling marked point patterns by intensity-marked Cox processes.
\newblock \emph{Statist. Probab. Lett.} 
\textbf{78}, 1194--1199.

\bibitem[{Illian \emph{et~al.}(2008)Illian, Penttinen, Stoyan and
  Stoyan}]{IllianEtal2008}
Illian, J., Penttinen, A., Stoyan, H. and Stoyan, D. (2008).
\newblock \emph{Statistical analysis and modelling of spatial point patterns}.
\newblock 
John Wiley \& Sons, Ltd, Chichester.

\bibitem[{Loosmore and Ford(2006)}]{LoosmoreFord2006}
Loosmore, N.~B. and Ford, E.~D. (2006).
\newblock Statistical inference using the G or K point pattern spatial
  statistics.
\newblock \emph{Ecology} \textbf{87}, 1925--1931.

\bibitem[{M{\o}ller and Berthelsen(2012)}]{MollerBerthelsen2012}
M{\o}ller, J. and Berthelsen, K.~K. (2012).
\newblock Transforming spatial point processes into Poisson processes using
  random superposition.
\newblock \emph{Adv. in Appl. Probab.} 
\textbf{44}, 42--62.

\bibitem[{Myllym{\"a}ki(2009)}]{Myllymaki2009}
Myllym{\"a}ki, M. (2009).
\newblock \emph{Statistical models and inference for spatial point patterns
  with intensity-dependent marks}.
\newblock {Ph.D.} thesis, University of Jyv{\"a}skyl{\"a}, Jyv{\"a}skyl{\"a}.

\bibitem[{Myllym\"aki and Penttinen(2009)}]{MyllymakiPenttinen2009}
Myllym\"aki, M. and Penttinen, A. (2009).
\newblock Conditionally heteroscedastic intensity-dependent marking of log
  {Gaussian Cox} processes.
\newblock \emph{Stat. Neerl.} 
\textbf{63}, 450--473.

\bibitem[{Penrose and Shcherbakov(2009)}]{PenroseShcherbakov2009}
Penrose, M.~D. and Shcherbakov, V. (2009).
\newblock Maximum likelihood estimation for cooperative sequential adsorption.
\newblock \emph{Adv. in Appl. Probab.} 
\textbf{41}, 978--1001.

\bibitem[{Penttinen and Stoyan(1989)}]{PenttinenStoyan1989}
Penttinen, A. and Stoyan, D. (1989).
\newblock Statistical analysis for a class of line segment processes.
\newblock \emph{Scand. J. Stat.} 
\textbf{16}, 153--168.

\bibitem[{Ripley(1976)}]{Ripley1976}
Ripley, B.~D. (1976).
\newblock The second-order analysis of stationary point processes.
\newblock \emph{J. Appl. Probab.} 
\textbf{13}, 255--266.

\bibitem[{Ripley(1977)}]{Ripley1977}
Ripley, B.~D. (1977).
\newblock Modelling spatial patterns.
\newblock \emph{J. R. Stat. Soc. Ser. B Stat. Methodol.} 
\textbf{39}, 172--212.

\bibitem[{Ripley(1979)}]{Ripley1979}
Ripley, B.~D. (1979).
\newblock Tests of 'randomness' for spatial point patterns.
\newblock \emph{J. R. Stat. Soc. Ser. B Stat. Methodol.} 
\textbf{41}, 368--374.

\bibitem[{Schladitz and Baddeley(2000)}]{SchladitzBaddeley2000}
Schladitz, K. and Baddeley, A.~J. (2000).
\newblock A third order point process characteristic.
\newblock \emph{Scand. J. Stat.} 
\textbf{27}, 657--671.

\bibitem[{Schlather(2001)}]{Schlather2001}
Schlather, M. (2001).
\newblock On the second-order characteristics of marked point processes.
\newblock \emph{Bernoulli} \textbf{7}, 99--117.

\bibitem[{Schlather \emph{et~al.}(2004)Schlather, Ribeiro and
  Diggle}]{Schlatheretal2004}
Schlather, M., Ribeiro Jr., P.~J. and Diggle, P.~J. (2004).
\newblock Detecting dependence between marks and locations of marked point
  processes.
\newblock \emph{J. R. Stat. Soc. Ser. B Stat. Methodol.} 
\textbf{66}, 79--93.

\bibitem[{Th{\"o}nnes and van Lieshout(1999)}]{ThonnesLieshout1999}
Th{\"o}nnes, E. and van Lieshout, M.-C. (1999).
\newblock A comparative study on the power of {van Lieshout} and {Baddeley's}
  {J--function}.
\newblock \emph{Biom. J.} 
\textbf{41}, 721--734.

\bibitem[{van Lieshout and Baddeley(1996)}]{LieshoutBaddeley1996}
van Lieshout, M. N.~M. and Baddeley, A.~J. (1996).
\newblock A nonparametric measure of spatial interaction in point patterns.
\newblock \emph{Stat. Neerl.} 
\textbf{50}, 344--361.

\end{thebibliography}

\section*{Appendix: The power in the toy examples}

\subsection*{A.1 Toy example 1}

Let $X_i \sim N(\mu_i,\sigma_i^2)$, $i=1,\dots,n$, $U_{\infty} = \max_{i}
|X_i|$ and $U_{\infty,w} = \max_{i} (w_i |X_i|)$ with $w_i=1/\sigma_i$.
Then $Y_i=|X_i|$ follows the folded normal distribution
with cumulative distribution function
\begin{eqnarray}\label{Toyexample_F_Yi}
F_{Y_i}(y; \mu_i, \sigma_i^2) = \Phi\left(\frac{y-\mu_i}{\sigma_i}\right) + \Phi\left(\frac{y+\mu_i}{\sigma_i}\right), \quad y\geq 0,
\end{eqnarray}
where $\Phi(\cdot)$ is the cumulative distribution function of the standard normal distribution.
Further, the distribution function of $U_{\infty}$ is
\begin{eqnarray}\label{Toyexample_F_Dinfty}
\nonumber F_{U_{\infty}}(u; \mu_1,\dots,\mu_n, \sigma_1,\dots,\sigma_n) &=& \mathbf{P}\left(\max_{i=1,\dots,n} Y_i \leq u\right)\\ &=& \prod_{i=1}^n P(Y_i \leq u)
= \prod_{i=1}^n F_{Y_i}(u; \mu_i, \sigma_i^2).
\end{eqnarray}
If $\mu_i=0$, the distribution \eqref{Toyexample_F_Yi} is called half-normal distribution and it simplifies to
$$
F_{Y_i}(y, 0, \sigma_i^2) = \int_0^y \frac{1}{\sigma_i}\sqrt{\frac{2}{\pi}}\exp\left(-\frac{u^2}{2\sigma_i^2}\right)\text{d}u.
$$
The critical value $c_{\text{crit}}$ for the null hypothesis that $\mu_i=0$ for all $i=1, \dots,n$ can be obtained by solving $F_{U_{\infty}}(c_{\text{crit}}; 0,\dots,0, \sigma_1,\dots,\sigma_n) = 1-\alpha$,
where $\alpha$ is the significance level and $F_{U_{\infty}}$ is given in \eqref{Toyexample_F_Dinfty}.
Thereafter the power of the unscaled test for the alternative hypothesis $H_1$: $\mu_i \neq 0$ for $i \in I_1$, where $I_1$ is a subset of $I= \{1,\dots,n\}$, can be obtained from
$$
\mathbf{P}\left(\max_{i=1,\dots,n} |X_i| > c_{\text{crit}}\right) = 1 - F_{U_{\infty}}(c_{\text{crit}}; \mu_1,\dots,\mu_n, \sigma_1,\dots,\sigma_n).
$$
The power of the scaled test can be obtained similarly, because for $Z_i = w_i Y_i = w_i|X_i|$ it holds
$$
F_{Z_i}(u;\mu_i,\sigma_i^2) = \mathbf{P}(Z_i \leq u) = \mathbf{P}(Y_i \leq u/w_i) = F_{Y_i}(u/w_i, \mu_i, \sigma_i^2)
$$
and the distribution of $U_{\infty,w}$ is
\begin{eqnarray}\label{Toyexample_F_Dinftyw}
F_{U_{\infty,w}}(u; \mu_1,\dots,\mu_n, \sigma_1,\dots,\sigma_n) = \mathbf{1}\left(\max_{i=1,\dots,n} Z_i \leq u\right) = \prod_{i=1}^n F_{Z_i}(u, \mu_i, \sigma_i^2).
\end{eqnarray}

\subsection*{A.2 Toy example 2}

Assume the random variable $X_i$ has the distribution \eqref{Toyexample2_FXi} and $Y_i = |X_i|$.
Since $\mathbf{P}(Y_i\leq y) = \mathbf{P}(-y\leq X_i\leq y)$,
the distribution of $Y_i$ is
\begin{eqnarray}\label{Toyexample2_FYi}
F_{Y_i}(y; \mu_i, \sigma_{ai}^2, \sigma_{bi}^2) &=& F_{X_i}(y; \mu_i, \sigma_{ai}^2, \sigma_{bi}^2) - F_{X_i}(-y; \mu_i, \sigma_{ai}^2, \sigma_{bi}^2) \\
\nonumber &=& \frac{1}{2} (F_1(y; \mu_i, \sigma_{ai}^2)-F_1(-y; \mu_i, \sigma_{ai}^2)) + \\
\nonumber && \frac{1}{2} (F_2(y; \mu_i, \sigma_{bi}^2)-F_2(-y; \mu_i, \sigma_{bi}^2)),
\end{eqnarray}
for $y>0$ (0 otherwise).
Then the distribution of $U_{\infty} = \max_{i} Y_i$ is
\begin{eqnarray*}
F_{U_{\infty}}(u; \mu_1,\dots,\mu_n, \{\sigma_{ai}, \sigma_{bi}\}) = \prod_{i=1}^n F_{Y_i}(u; \mu_i, \sigma_{ai}^2, \sigma_{bi}^2),
\end{eqnarray*}
similarly as \eqref{Toyexample_F_Dinfty}.
Let then $Z_i= \mathbf{1}(X_i \geq 0) w_i^{(+)} X_i +  \mathbf{1}(X_i<0) w_i^{(-)} X_i$, where
$w_i^{(+)}$ and $w_i^{(-)}$ are weights.
The distribution of $Z_i$ is
\begin{eqnarray*}
F_{Z_i}(u; \mu_i, \sigma_{ai}^2, \sigma_{bi}^2) &=& \mathbf{P}(Z_i \leq u) \\ &=&
\mathbf{P}\left(Z_i \leq u | X_i\geq 0\right)\mathbf{P}\left(X_i\geq 0\right) + \mathbf{P}\left(Z_i \leq u | X_i < 0\right)\mathbf{P}\left(X_i< 0\right) \\
&=& \mathbf{P}\left(Z_i \leq u  \; \& \; X_i\geq 0\right) + \mathbf{P}\left(Z_i \leq u  \; \& \; X_i < 0\right) \\
&=& \mathbf{P}\left(0 \leq X_i \leq u/ w_i^{(+)}\right) + \mathbf{P}\left(-u/ w_i^{(-)} \leq X_i \leq 0\right) \\
&=& F_{X_i}\left(u/ w_i^{(+)}; \mu_i, \sigma_{ai}^2, \sigma_{bi}^2\right) - F_{X_i}\left(-u/ w_i^{(-)}; \mu_i, \sigma_{ai}^2, \sigma_{bi}^2\right),
\end{eqnarray*}
and the distribution of $U_{\infty,w} = \max_{i} Z_i$ is obtained similarly as \eqref{Toyexample_F_Dinftyw}:
\begin{eqnarray*}
F_{U_{\infty,w}}(u; \mu_1,\dots,\mu_n, \{\sigma_{ai}, \sigma_{bi}\}) = \prod_{i=1}^n F_{Z_i}(u; \mu_i, \sigma_{ai}^2, \sigma_{bi}^2).
\end{eqnarray*}

The powers of the tests based on $U_{\infty}$ and $U_{\infty,w}$
can then be calculated in the same way as in the case of the Toy example 1 above.
That is, for $U_{\infty,w}$, solve first the critical value $c_{\text{crit}}$ for
the null hypothesis that $\mu_i=0$ for all $i$ from
$F_{U_{\infty,w}}(c_{\text{crit}}; 0,\dots,0, \{\sigma_{ai}, \sigma_{bi}\}) = 1-\alpha$,
and then calculate
$\mathbf{P}\left(\max_{i=1,\dots,n} Z_i > c_{\text{crit}}\right) = 1 - F_{U_{\infty},w}(c_{\text{crit}}; \mu_1,\dots,\mu_n, \{\sigma_{ai}, \sigma_{bi}\})$
for the alternative hypothesis with $\mu_1,\dots,\mu_n$ where $\mu_i \neq 0$ for $i \in I_1\subseteq I$,
and similarly for $U_{\infty}$.

\end{document}